# Identity Theft in AI Conference Peer Review

Nihar B. Shah[1], Melisa Bok[2], Xukun Liu[2], Andrew McCallum[2,3]
[1]Carnegie Mellon University and OpenReview.net board
[2]OpenReview.net
[3]University of Massachusetts

**Abstract**: We discuss newly uncovered cases of identity theft in the scientific peer-review process within artificial intelligence (AI) research, with broader implications for other academic procedures. We detail how dishonest researchers exploit the peer-review system by creating fraudulent reviewer profiles to manipulate paper evaluations, leveraging weaknesses in reviewer recruitment workflows and identity verification processes. The findings highlight the critical need for stronger safeguards against identity theft in peer review and academia at large, and to this end, we also propose mitigating strategies.

Academia heavily relies on trust. This trust-based system, however, creates a significant vulnerability: identity theft. In this report, we describe newly uncovered cases of identity theft within the scientific peer-review process within the research area of artificial intelligence (AI), involving modus operandi that could also disrupt other academic procedures.

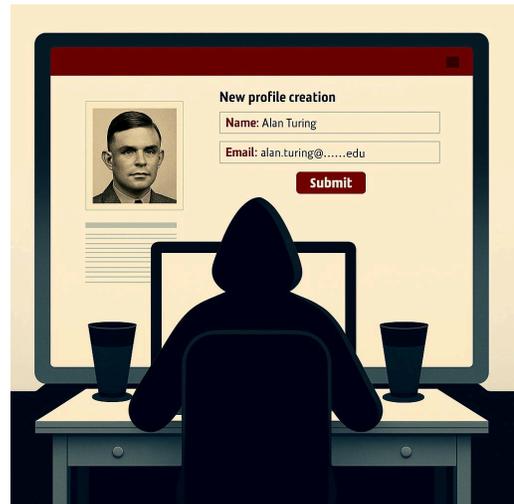

To set the stage for the discussions that follow, we begin by outlining the peer-review process, focussing on scientific conferences since they are the most prominent venues of publication in computer science. Peer review is foundational to scientific inquiry, relying on researchers to voluntarily apply their expertise in evaluating scientific papers. As the need for large and diverse reviewing pools has grown, the reviewer-recruitment process has increasingly included open calls where researchers sign up using online forms. The online signup forms typically ask for details about the researcher's designation, institutional affiliation, links to prior publications, and contact information. These procedures are common not only in AI and other subfields of computer science such as HCI and security, but also in disciplines like astronomy and biomedicine. Other means of reviewer recruitment include profiles created by the potential reviewers themselves on online platforms, or potential reviewers emailing program chairs offering to volunteer reviewing services. In each of these situations, after determining that an applicant has the necessary expertise, program chairs enlist them in the review process using the provided contact information.

After recruitment, reviewers are assigned to evaluate submitted papers that align with their expertise. This assignment of reviewers to papers is typically performed by (semi-)automated systems. Reviewers express their interest and expertise in reviewing specific submissions through a process known as "bidding," as well as by outlining their qualifications in a "reviewer profile" comprising their publication record. These elements are then fed into an algorithm that matches reviewers to papers (see [9, Section 3], for more details). Once assigned, reviewers are expected to provide objective and unbiased evaluations of the submissions. The outcomes of these reviews—acceptances or rejections of the submitted papers—carry substantial consequences for the authors' professional trajectories.

In computer science, the most significant demand for large and rapidly expanding reviewer pools is found in AI conferences which accept roughly 15-25% of papers submitted to them. Rising interest in AI is straining its peer review process—some AI conferences receive thousands or even tens of thousands of paper submissions at their annual deadlines, having grown nearly tenfold in the past ten years. They also have a comparable number of reviewers needing to be assigned within just a few days. Currently, obtaining a verified, internationally recognized proof of identity for reviewers online is practically unfeasible. For example, platforms like ORCID, which provide researcher IDs, do not verify personal identities, nor prevent duplicates. As a result, reliably verifying the large influx of accounts is a major, unsolved challenge.

OpenReview.net is a nonprofit platform that provides free, flexible, and scalable infrastructure for managing peer-review workflows, supporting over 2,000 conferences and workshops, including most top-tier AI conferences. In this manuscript we report attempts to manipulate the peer-review process, which were identified and investigated by OpenReview staff. Similar issues may affect other platforms with comparable workflows; while many peer-review venues, including those outside AI, use open forms for reviewer (self-)nomination, we are not aware of any similar investigations elsewhere. We report these findings in the spirit of transparency, aiming to alert the broader academic community. It is our hope that by sharing these insights, other peer-review platforms and peer-review venues can improve vigilance against such manipulations.

**Findings of fraud**

Our findings are based on in-depth investigations conducted in the context of multiple AI conferences, between February and April 2024 and then in April 2025, requiring extensive time from OpenReview staff. This investigation unearthed 94 reviewer (and meta-reviewer) profiles involving fake identities. Under such a fraud scheme, the dishonest researcher employs the following modus operandi to obtain more favorable reviews for papers authored by their true identity.

- If reviewers are recruited through an online form, a dishonest researcher submits the form using another researcher's identity, including the other researcher's affiliations and publication history. In the form, the dishonest researcher provides an email address that the dishonest researcher controls.
- If recruitment through an online form is not provided as an option, but the program chairs recruit through profiles on review platforms, the dishonest researcher creates a profile with someone else's affiliation and publication history.
- The program chairs verify that, based on the supplied information about seniority and/or prior publication history, this (fraudulent) reviewer is eligible to be a reviewer.
- Once embedded in the review system, the dishonest researcher under this fraudulent identity attempts to get assigned to review papers authored by their true identity. This can sometimes be accomplished by expressing interest in reviewing the paper during bidding [6]. This may alternatively be accomplished by increasing their perceived suitability as a reviewer for the paper by carefully tuning the false reviewer profile or some text in their own paper [5].
- Once assigned, the dishonest researcher—as the reviewer under their false identity—provides favorable reviews to the papers authored by their true identity.

We have also found researchers creating not just one but multiple fake reviewer profiles in order to ensure a higher chance of success in achieving their fraudulent goals. Another variant is where a dishonest researcher may indulge in a collusion ring, where they team up with another dishonest researcher to favorably review their co-conspirators' work in exchange for some quid-pro-quo.

One prospective remedy is to mandate the use of institutional email addresses as a pseudo-identity, and verify that the provided affiliations correspond with the domain of the email. However, this approach has multiple problems: a surprising number of appropriate users don't have an institutional email, others strongly prefer not to use one, and there is no clear categorization of which domains should be considered trusted institutions. Moreover, even highly trusted domains can be exploited. In all 94 cases, the fake reviewer profiles included a round-trip-verified email, meaning OpenReview sent a verification email to the address, and the email holder confirmed control by clicking the verification link. Of these, 92 email addresses pertained to reputed universities, while two had ".edu" domains of defunct institutions. In many of these cases, we found that the dishonest researcher had leveraged the current common practice in which universities allow members and visitors to create email aliases:

- The dishonest researcher gained access to an email of a trusted institution, and created email alias(es) resembling someone else at that institution. Alternatively, the dishonest researcher may ask a co-conspirator at another institution to create such an alias.
- The dishonest researcher then signed up to review using the other person's identity and the aforementioned email address, by employing the modus operandi outlined earlier.

This bypasses further checks on whether the email pertains to the domain of the claimed affiliation.

Finally, many conferences automatically import reviewer lists from previous years, creating a vulnerability where undetected impersonators could be perpetually invited to review future editions.

The discovery of these fraud cases adds to related issues of misconduct in peer review [1, 9 (Section 4)] including collusion rings in computer science [7, 8, 10] and similar issues in other fields [2, 3, 4]. Given the numerous ways that the reviewer-paper assignment is successfully manipulated, our community should actively explore protective measures. Some venues have introduced randomness in the assignment process [6] which limits the chances of being assigned a colluding reviewer, but stronger solutions are needed.

Peer-review venues and platforms such as OpenReview.net are increasingly investigating and catching inappropriate behavior, before, during, and after the review process. Consequences for the misbehavior can be severe, such as prohibition from publishing or being reported to the dishonest researcher's employer. However, it is preferable to proactively prevent such behavior rather than addressing it after the fact.

**Combating such fraud**

In the longer term, our community should devise improved identity verification procedures through social and cryptographic means. We call upon the scientific community to pay greater attention to these problems that threaten the integrity of science. Based on our observations, we make the following recommendations to program chairs, editors, and peer-review management platforms. Several of these recommendations have already been implemented on the OpenReview platform.
- A credential, though not as strong as a "real ID," is the reviewer's previously published papers, which must be verified to ensure they were authored by the individual claiming to be the reviewer. One approach to this is to verify that both the prior publications and the current review were submitted through the same email address, or the same account when the same peer-review platform is used for both.

- A second approach to publication linkage is possible when a more general persistent digital identifier such as ORCID is employed. The peer-review platform can then require the reviewer to log in through the identifier system to verify a match between the reviewer's claimed ID and the ID associated with previously published papers. This would require a greater integration of ORCID in the computer science publishing process. Similarly, any verified author profiles and associated publication history published through ACM digital library or IEEE Xplore may be used.
- For reviewers signing up on a system where they haven't previously published, implement carefully designed vouching protocols to deter misconduct, provided appropriate network-propagating consequences are in place. For instance, the preprint platform arXiv requires vouching of new authors (from unverified institutions) by another researcher who has previously published on arXiv.
- Pay particular attention to reviewer profiles created shortly before the conference deadline to identify any potential issues.
- Public visibility of the created researcher profiles on many platforms offers an opportunity for the broader scientific community to contribute to some of these efforts.
- Traditional anti-fraud techniques such as those using IP addresses or browser fingerprints may also be considered.
- Platforms that manage researcher profiles should implement strong, regularly-enforced deduplication methods. Stronger deduplication can also help prevent scenarios of individuals creating multiple profiles with slight variations (e.g., different affiliations) to bypass conflict of interest rules.
- Make reviewer-paper assignment processes more robust to gamification [5, 6, 9 (Section 4)].

We also make the following recommendations to universities and organizations with email domains. We believe these steps are also in their best interest since it mitigates the risk of fraud using the organization's email.
- Implement stricter monitoring of alias email accounts created under their domains. For instance, if someone who is *not* Alan Turing creates an email address alan.turing@... then this should raise a red flag.
- To confirm the legitimacy of email addresses, allow third-party verification of email ownership.
- Conduct thorough investigations when cases of identity theft are reported.

We encourage making public the investigations and (potentially anonymized) outcomes of scientific misconduct in order to foster transparency, inform other organizations, and enable research on mitigation strategies.

Finally, other critical academic processes may be similarly vulnerable. Applications for university admissions, internships, and fellowships often depend on recommendation letters, where the applicant supplies the recommender's contact information, leaving room for abuse through falsified identities and unverified email addresses. Moreover, the misuse of email aliases for impersonation, as uncovered in our investigation, could be exploited beyond academic workflows, including for phishing or other forms of deception. These findings highlight the need for heightened vigilance across academic systems. Sustained, transparent discourse about such vulnerabilities is essential to safeguarding the integrity of our shared scientific processes.

**References**

[1] Aiken, A., Amato, N.M., Bowling, K., De Floriani, L., de Sturler, E., Gini, M., Hanson, V., Krishnamurthy, A., Larson, K., Li, W., Littman, M., Ozcan, F., Russell, M., Sarkar, V., Schwartz, A., Spafford, E.H., and Srivastava, D., Report of the CRA Working Group on Research Integrity. August 2023.

[2] Cohen, A., Pattanaik, S., Kumar, P., Bies, R.R., De Boer, A., Ferro, A., Gilchrist, A., Isbister, G.K., Ross, S. and Webb, A.J., 2016. Organised crime against the academic peer review system. British Journal of Clinical Pharmacology, 81(6), p.1012.

[3] Dadkhah, M., Lagzian, M. and Borchardt, G., 2018. Identity theft in the academic world leads to junk science. Science and Engineering Ethics, 24, pp.287-290.

[4] Ferguson, C., Marcus, A. and Oransky, I., 2014. The peer-review scam. Nature, 515(7528), p.480.

[5] Hsieh J., Raghunathan A., Shah, N., 2024. Vulnerability of Text-Matching in ML/AI Conference Reviewer Assignments to Collusions. arXiv:2412.06606.

[6] Jecmen, S., Zhang, H., Liu, R., Shah, N., Conitzer, V. and Fang, F., 2020. Mitigating manipulation in peer review via randomized reviewer assignments. Advances in Neural Information Processing Systems, 33, pp.12533-12545.

[7] Littman M. Collusion rings threaten the integrity of computer science research. Communications of the ACM. 2021 May 24;64(6):43-4.

[8] Rastogi, C., Song, X., Jin, Z., Stelmakh, I., Daumé III, H., Zhang, K., and Shah, N, 2024. A Randomized Controlled Trial on Anonymizing Reviewers to Each Other in Peer Review Discussions. PLOS ONE. Dec 2024.

[9] Shah N. Challenges, Experiments, and Computational Solutions in Peer Review (Extended Version). Available online: https://www.cs.cmu.edu/~nihars/preprints/SurveyPeerReview.pdf. Shorter version published in the Communications of the ACM, 65(6), pp.76-87.

[10] T. N. Vijaykumar. Potential Organized Fraud in On-Going ASPLOS Reviews. Blog: https://medium.com/@tnvijayk/potential-organized-fraud-in-on-going-asplos-reviews-874ce14a3ebe. Nov 2020.